\documentclass[final, runningheads]{llncs}
\usepackage[T1]{fontenc}
\usepackage{graphicx}
\usepackage{amsmath}
\newcommand{\itadata}{\footnotesize \textsl{Workshop Scientific HPC in the pre-Exascale era (part of ITADATA2024)}}
\usepackage{fancyhdr}
\fancypagestyle{empty}{\fancyhf{}\fancyhead[C]{\itadata}
  \fancyhead[R]{\includegraphics[width=.05\textwidth]{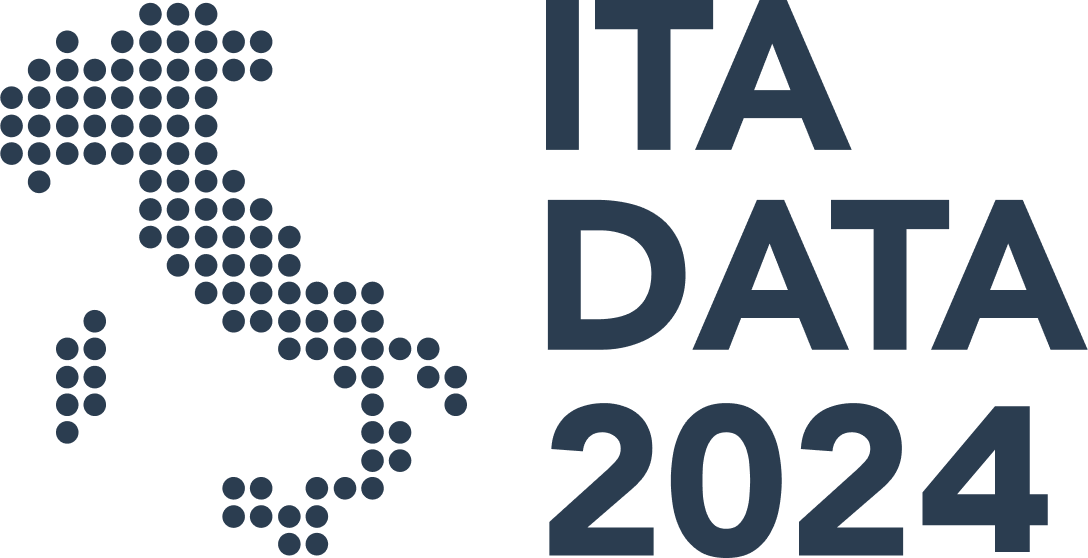}}}

\makeatletter
\begin{document}
\title{The Problem of the Global Astrometric Sphere Reconstruction in Astrometry - Issues and Approaches}
\author{Alberto Vecchiato\inst{1}\orcidID{0000-0003-1399-5556} \and
Alexey Butkevich\inst{1}\orcidID{0000-0002-4098-3588} \and
Mario Gai\inst{3}\orcidID{0000-0001-9008-134X} \and
Valentina Cesare\inst{1,2}\orcidID{0000-0003-1119-4237} \and
Beatrice Bucciarelli\inst{1}\orcidID{0000-0002-5303-0268} \and
Mario G. Lattanzi\inst{1}\orcidID{0000-0003-0429-7748}}
\authorrunning{A. Vecchiato et al.}
\institute{National Institute of Astrophysics - Astrophysical Observatory of Torino, via Osservatorio 20, 10025 Pino Torinese, Italy \\
\email{alberto.vecchiato@inaf.it} \and
National Institute of Astrophysics - Astrophysical Observatory of Catania, Via Santa Sofia 78, 95123 Catania, Italy
}
\maketitle              \begin{abstract}
In this contribution we give a brief account of the problem of the Global Astrometric Sphere Reconstruction in Astrometry, with particular reference to the Gaia and Gaia-like astrometric missions, namely those adopting a scanning strategy with observations in TDI mode.  We sketch the design of the Gaia mission, the mathematical modelling that comes naturally from its observing strategy, and how the problem of the global sphere reconstruction translates into that of the solution of large, sparse, and overdetermined system of linearized equations. After a short description of the two approaches to this problem implemented in the Gaia data reduction pipelines, we list the main known problems of the current approaches, with specific reference to the calibration and the correlation issues. Finally, we suggest how an arc-based solution could help to alleviate some of these problems, how it would be possible to devise a mathematical model for such an observable despite the TDI observing mode, and the main difficulty that a parallel implementation of this model would have to solve.

\end{abstract}
\section{What is Astrometry About}
Astrometry is a branch of Astronomy that deals with the estimation of the position and motion of sources on the sky, by angle measurements with respect to other sources or to the local reference frame of an instrument. Astrometry can in turn be \emph{relative} or \emph{global}, the latter technique applying when the goal is the definition of a reference system for the whole celestial sphere \cite{2002moas.book.....K}. Dating back to several thousands years ago, it is one of the most ancient scientific disciplines, and after more than a century of decline caused by the expansion of the astronomical applications, it has regained a greater influence thanks to the technological advances that allowed the advent of global astrometry from space, first with the Hipparcos satellite, and than with its successor Gaia.

\section{The Gaia Mission}
Gaia is a satellite and a mission of the European Space Agency (ESA) launched in December 2013. Its main goal is the production of a five-parameter astrometric catalog (i.e., including positions, parallaxes and the two components of the proper motions) at the 10 to 1000~$\mu$arcsecond-level ($\mu$as) for about one billion stars of our Galaxy in the magnitude range from 3 to 20.7 \cite{2016A&A...595A...1G}. The initial five-year duration of the operational phase has more than doubled, and the mission is expected to end at the beginning of 2025 due to the depletion of the fuel required for operations.

Currently, the Data Processing and Analysis Consortium (DPAC) in charge of processing the satellite data has produced three Data Releases (DR), respectively in 2016, 2018, and 2020-2022. The DR4 is planned to be published after mid 2026, while the DR5, the final release based on all the mission data, not before the end of 2030.

\begin{figure}
    \begin{center}
        \includegraphics[width=0.6\textwidth]{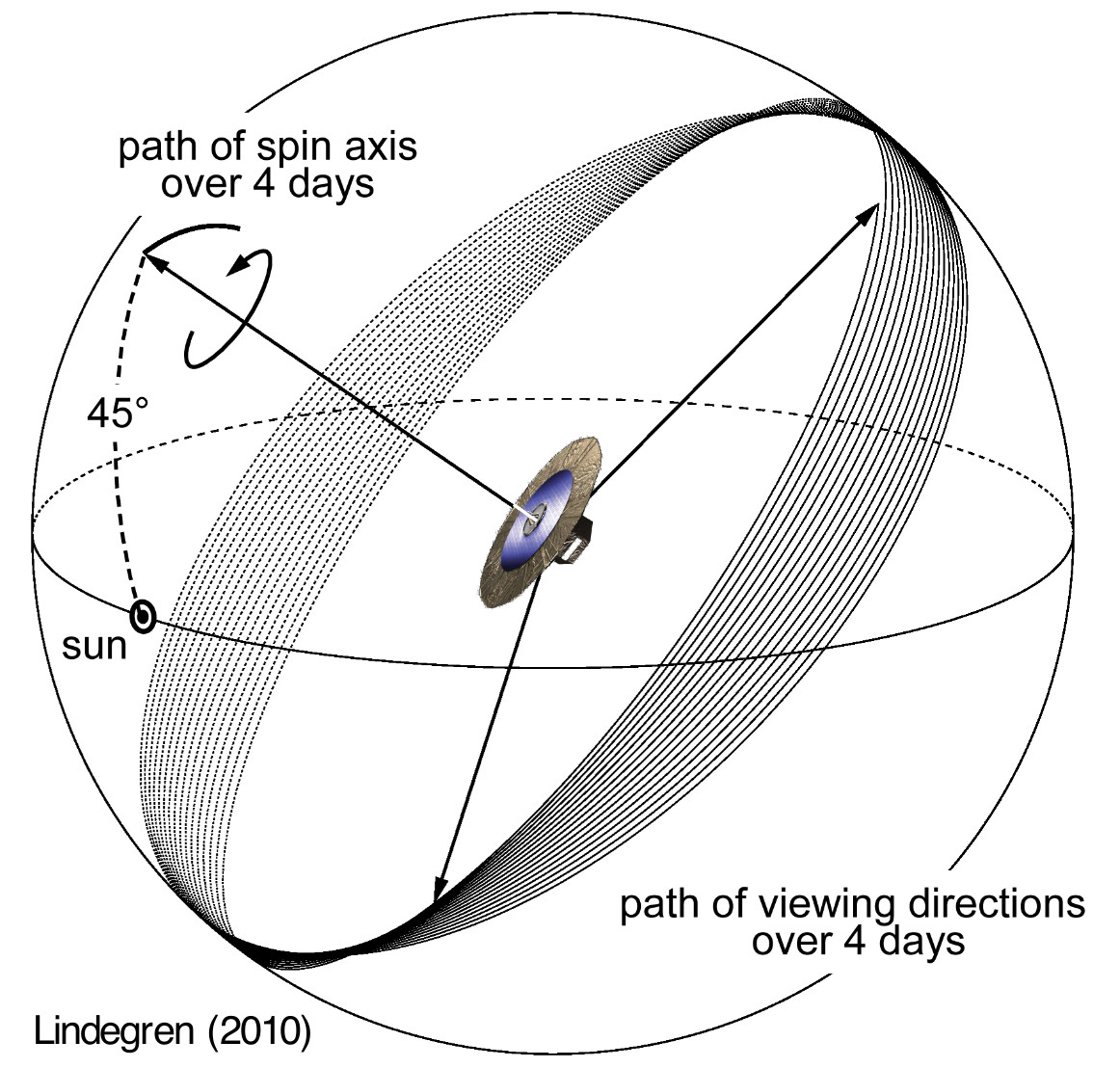}
    \end{center}
    \caption{Schematic representation of the Gaia Nominal Scanning Law.} \label{FigNSL}
\end{figure}

The mission is designed as a scanning satellite that continuously sweeps the sky following a predetermined Nominal Scanning Law (NSL) obtained from the combination of three independent motions, namely a spin around the satellite axis, a precession of the latter around the Sun-satellite direction, and is orbital motion around the Sun (see Fig.~\ref{FigNSL}). The NSL induces a complete coverage of the celestial sphere every six months, which implies that the sky is observed repeatedly (about 20 times) over the entire mission duration.

\begin{figure}
    \begin{center}
        \includegraphics[width=\textwidth]{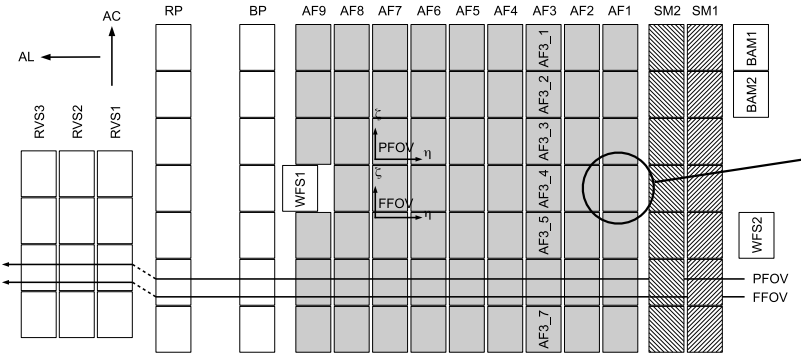}
    \end{center}
    \caption{The focal plane of the telescope is covered by a mosaic of 106 CCDs, 62 of which constitute the part devoted to collect the high-precision astrometric data (AF1 to AF9). The projection of the sources sweep the entire focal plane mainly along the AL direction in about 40 seconds.} \label{FigFocPlane}
\end{figure}

The satellite observes the sky through two different Fields of View (FoV) whose pointing directions are separated by a Basic Angle (BA) of about $106.5^\circ$, which helps to measure the absolute parallaxes, and therefore the distances, of the objects. The light from the observed sources are conveyed to the focal plane of the telescope, which is covered by a mosaic of CCDs (Fig.~\ref{FigFocPlane}). Because of the scanning law, the light of the sources sweeps along the CCDs, and the collected photoelectrons are transported in-sync with the satellite movement; such operating mode is called ``Time-Delay and Integration'' (TDI), and allows to collect all the photoelectrons coming from a specific source when they cross a given fiducial line of the CCD. The crossing time, and possibly its ``vertical'' position, is also recorded. This data constitute the essential information that is needed by the Global Astrometric Sphere Reconstruction process.

\section{Global Astrometric Sphere Reconstruction with Gaia}
\begin{figure}
    \begin{center}
        \includegraphics[width=\textwidth]{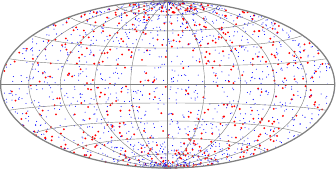}
    \end{center}
    \caption{Ideal representation of the primary stars (in red) and of the attached secondary stars (in blue).} \label{FigPriSec}
\end{figure}

At the core of the realization of the complete catalog is a process, called ``Global Astrometric Sphere Reconstruction,'' whose goal is the definition of a global celestial reference frame linked to the Barycentric Celestial Reference System (BCRS, \cite{2005USNOC.179.....K}). First, this procedure fixes the positions, parallaxes, and proper motions of a subset of $\sim10^{7}\text{--}10^{8}$ ``primary stars,'' namely those sources that are sufficiently ``well-behaved'' to be modelled by these 5 standard astrometric parameters. This phase includes also the refinement of the instrument calibration needed to achieve the required catalog accuracy, as well as the reconstruction of the satellite attitude, and provides the first materialization of such a frame of reference. The remaining stellar objects (secondary stars), which include, e.g., multiple stars with too short a period, or stars with a variability too large, can then be reduced by considering only their astrometric parameters as unknowns. The secondary sources' observations are reduced in the reference system previously defined by the primaries, and therefore concur to densify the reference frame (Fig.~\ref{FigPriSec}). This complete two-steps process is realized by a pipeline called Astrometric Global Iterative Solution (AGIS; \cite{2012A&A...538A..78L}).

\begin{figure}
    \begin{center}
        \includegraphics[width=0.6\textwidth]{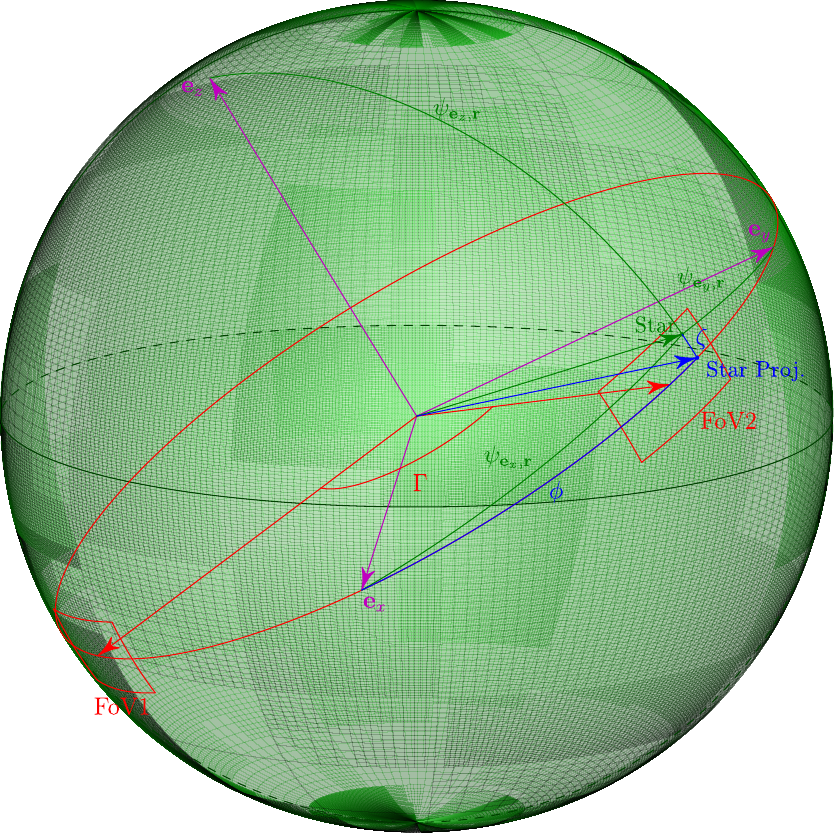}
    \end{center}
    \caption{Ideal representation of the primary stars (in red) and of the attached secondary stars (in blue).} \label{FigGaiaMeas}
\end{figure}

In the mathematical model used for the sphere reconstruction process, the above mentioned time of passage of a sources across a specific fiducial line of a single CCD is converted into an angle $\phi$ between the $x$ axis of the reference system of the satellite, defined as the direction bisecting the BA $\Gamma$, and the projection of the source position on the instantaneous scanning plane. This angle can be expressed as function of the direction cosines of the source vector $\mathbf{r}$ with respect to the axes of the satellite reference system $\mathbf{e}_{\hat{a}}$, $a=(x,y,z)$ (Fig.\ref{FigGaiaMeas}):
\begin{equation}
	\cos\psi_{(\hat{a},\mathbf{r})}=\frac{\mathbf{e}_{\hat{a}}\cdot\mathbf{r}}{\left|\mathbf{r}\right|}\label{eq:eucl-cosdir}
\end{equation}
\begin{equation}
    \cos\phi=\frac{\cos\psi_{(\hat{x},\mathbf{r})}}{\sqrt{1-\cos^{2}\psi_{(\hat{z},\mathbf{r})}}}=
    F\left(\boldsymbol{x}^{\mathrm{S}},\boldsymbol{x}^{\mathrm{A}},\boldsymbol{x}^{\mathrm{C}},\boldsymbol{x}^{\mathrm{G}}\right),
    \label{eq:abscissa}
\end{equation}
where $\boldsymbol{x}^{\mathrm{S},\mathrm{A},\mathrm{C},\mathrm{G}}$ identify the Source, Attitude, Calibration, and Global unknowns of the problem that are used for the primary sources. The source parameters are the sought for positions, parallaxes, and proper motions; the attitude parameters are those that define the orientation of the satellite reference system (Attitude) with respect to the BCRS; the calibration parameters are needed to refine the instrument calibration at the level needed by the target catalog accuracy; finally, there are a few global unknowns needed for special uses. A similar equation can be obtained for the Across Scan measurement ($\zeta$) when needed.

In this way, each measurement is converted into a non linear equation dependent on some of the unknowns of the global sphere reconstruction problem, which thus requires the solution of a large and sparse equation system, which is hardly possible. However, an approximate value of the parameters is already known, therefore the above equations can be linearized around these $\boldsymbol{x}_0$ values
\begin{align}
    -\sin\phi_{\mathrm{calc}\,\delta\phi} = &\sum_{\mathrm{Source}}\left.\frac{\partial F\left(\mathbf{x}\right)}{\partial\mathbf{x}^{\mathrm{S}}}\right|_{\mathbf{x}_{0}}\delta\mathbf{x}^{\mathrm{S}}+\sum_{\mathrm{Attitude}}\left.\frac{\partial F\left(\mathbf{x}\right)}{\partial\mathbf{x}^{\mathrm{A}}}\right|_{\mathbf{x}_{0}}\delta\mathbf{x}^{\mathrm{A}}\nonumber \\
     &+\sum_{\mathrm{Cal}}\left.\frac{\partial F\left(\mathbf{x}\right)}{\partial\mathbf{x}^{\mathrm{C}}}\right|_{\mathbf{x}_{0}}\delta\mathbf{x}^{\mathrm{C}}+\sum_{\mathrm{Global}}\left.\frac{\partial F\left(\mathbf{x}\right)}{\partial\mathbf{x}^{\mathrm{G}}}\right|_{\mathbf{x}_{0}}\delta\mathbf{x}^{\mathrm{G}},
\end{align}
where, if $\phi_{\mathrm{obs}}$ is the converted measurement of Gaia,
\begin{align*}
	\delta\phi & =\phi_{\mathrm{obs}}-\phi_{\mathrm{calc}}\\
	\phi_{\mathrm{calc}} & =F\left(\mathbf{x}_{0}\right),
\end{align*}
and if $\mathbf{x}_{\mathrm{true}}$ are the true values of the unknowns, the new unknowns of the problem are $\delta\mathbf{x}=\mathbf{x}_{\mathrm{true}}-\mathbf{x}_{0}$.

\begin{figure}
    \begin{center}
        \includegraphics[width=0.4\textwidth]{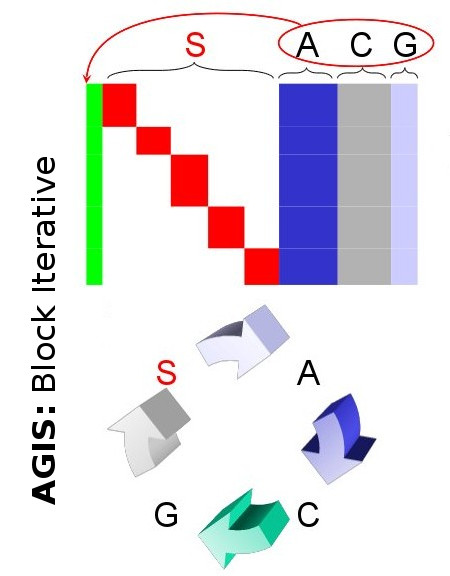}\hfill\includegraphics[width=0.4\textwidth]{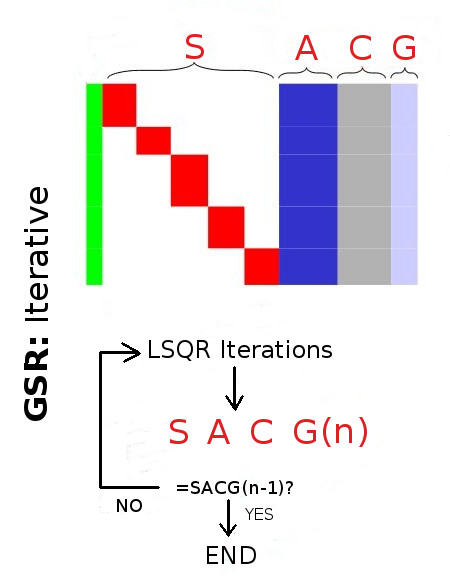}
    \end{center}
    \caption{The global sphere reconstruction in the AGIS pipeline (left) is implemented with a block-iterative algorithm. The GSR pipeline (right) uses a customized version of the fully iterative LSQR algorithm.} \label{FigAGISGSR}
\end{figure}

Since the number of observations $m\sim10^9-10^{10}$ is much larger than the number of unknowns $n\simeq5\times10^8$, and since each equation depends only on a few tens of unknowns, the problem of the sphere reconstruction is translated into the solution of a large, sparse, and overdetermined system of linear equations $\mathbf{b}=A\mathbf{x}$. The solution of this system $\mathbf{x}=\left(A^{T}A\right)^{-1}A^{T}\mathbf{b}$ requires the calculation of the pseudo-inverse matrix $\left(A^{T}A\right)^{-1}$. The sheer size of $n$ makes thus unfeasible the use of direct methods, which are $O(n^3)$ and would thus require $\sim10^{26}$ FLOP. One is therefore forced to resort to parallelized iterative algorithms. In Gaia the AGIS pipeline adopts a block iterative algorithm, which solves separately each class of unknowns (S, A, C or G). This allows for an embarassingly parallel implementation of the algorithm, but it makes much more difficult the estimation of the covariances between different classes of unknowns, which can bring to underestimating the parameters' errors. A verification pipeline called Global Sphere Reconstruction (GSR, \cite{2018A&A...620A..40V}) uses instead a customized version of the well-known LSQR iterative algorithm \cite{PaigeSaunders1982} (Fig.~\ref{FigAGISGSR}). At the price of a more complicated parallelization, this algorithm solves the whole system at the same time, and in principle allows the estimation of any element of the variance-covariance matrix \cite{2024SPIE13101E..1OC}.

\section{Known Issues of the Gaia Astrometric Catalog, their Origins, and a possible alternative approach to the sphere reconstruction}
The Gaia catalogs have experienced a number of issues like, e.g.:
\begin{itemize}
    \item A parallax bias of QSOs (zero-point parallax error) of $-17~\mu$as \cite{2021A&A...649A...2L}.
    \item Magnitude-, color-, and position-dependent parallax biases \cite{2021A&A...649A...4L}.
    \item Measurements' residuals not following the expected magnitude dependence.
\end{itemize}

The most probable main causes of these issues are a suboptimal calibration model, and the strong correlations that exist among different kind of unknowns. Both issues have been addressed through the various Gaia DRs, with significant improvement from DR2 to DR3. It is also expected that the situation will keep improving until the final release, however, given the non-trivial nature of the problem, it also worth listing the most significant correlations:
\begin{enumerate}
    \item Correlation between attitude and Basic Angle variations.
    \item Correlation between parallaxes and Basic Angle variations.
    \item Correlation between parallaxes and relativistic light deflection.
    \item Correlation between Basic Angle variations and other calibration parameters.
\end{enumerate}

\begin{figure}
    \begin{center}
        \includegraphics[width=0.6\textwidth]{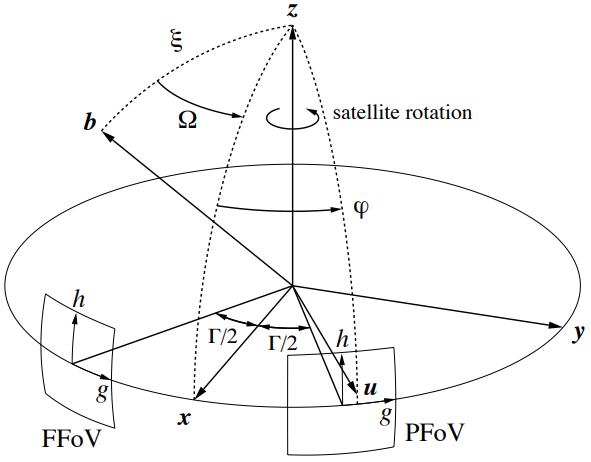}
    \end{center}
    \caption{An asymmetric variation of the orientation of the FoVs can be mimicked by a rotation of the satellite attitude around the $z$ axis.} \label{FigAttBACorr}
\end{figure}

The first correlation, in particular, depends on the purely geometric definition of the satellite's attitude where, as previously mentioned, the $x$ axis of the satellite is by definition the direction bisecting those of the two FoVs. For this reason, e.g., a rotation around the $z$ axis of one of the FoVs is equivalent to a rotation around the same axis of the $x$ and $y$ attitude axes (Fig.~\ref{FigAttBACorr}). The variations of the BA are measured by a dedicated instrument called Basic Angle Monitoring (BAM; \cite{2014RMxAC..45...35R}), so this issue might be considered benign as long as one combines the two single variations into a single change of the angle $\Gamma$.

\begin{figure}
    \begin{center}
        \includegraphics[width=0.6\textwidth]{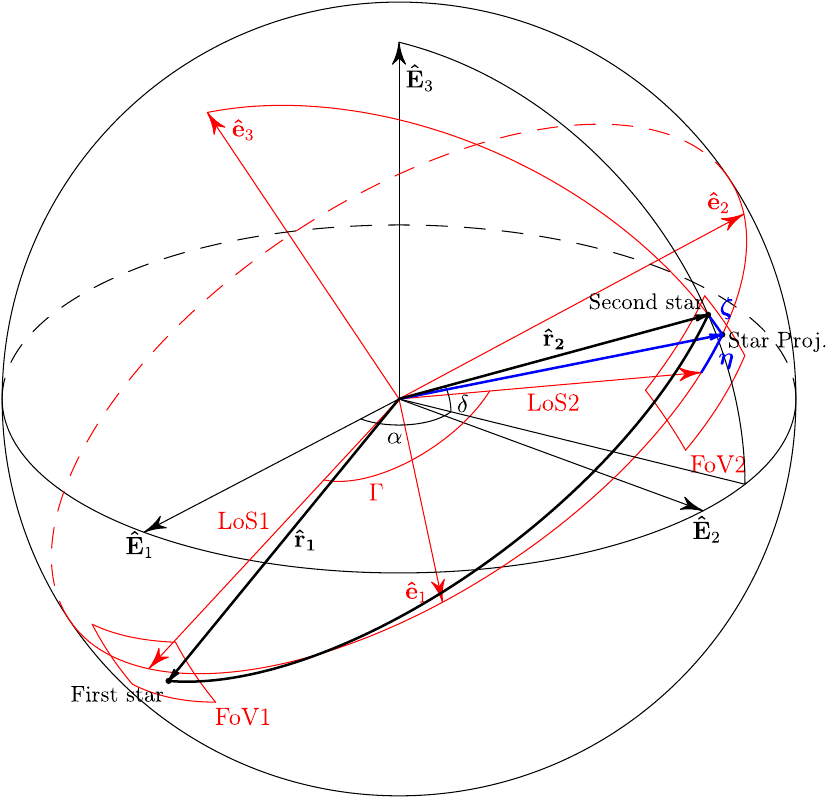}
    \end{center}
    \caption{Schematic representation of the elementary measurement of an arc-based model.} \label{FigArcBasedMmodel}
\end{figure}

On the other side, however, it has to be remembered that the attitude reconstruction requires some $\sim10^7$ unknowns, which constitute the second largest class of parameters of the equation system. Moreover, other kind of correlations, like the last one of the above list, might combine with the attitude, for example increasing the covariances and therefore degrading the accuracy of the final catalog. A possible way out of these difficulties is to change the mathematical model of the observable, switching from the one that uses the direction cosines of one single source with respect to the satellite reference system, to one that models directly the arc between two sources, namely
\begin{equation}
    \cos\psi=\frac{\mathbf{r}_{1}\cdot\mathbf{r}_{2}}{\left|\mathbf{r}_{1}\right|\left|\mathbf{r}_{2}\right|},
\end{equation}
where $\mathbf{r}_{1}$ and $\mathbf{r}_{2}$ and the vectors pointing to sources in different FoVs (Fig.~\ref{FigArcBasedMmodel}).

One of the difficulties that a model like this has to face is that it does not fit easily within the scanning approach of a Gaia-like mission. The definition of an arc, in fact, requires that the measurement of the two stars are given at the same time, which is intrinsically impossible from the very nature of the TDI mode. However this issue can be tackled by substituting the estimation of the attitude with the estimation of the so-called ``rates'', namely the angular velocities $\omega_\eta$ and $\omega_\zeta$, respectively along and across the scanning direction. These depend on the derivatives of the attitude, and while the attitude has to be determined continuously for the whole mission, the rates can be estimated for a limited time range, separately from the sphere reconstruction.

Another more HPC-related issue is that an arc-based model requires an extra effort for the parallelization of the algorithm that solves the linearized system of equations. Indeed, in the original model based on the attitude each observation contains the astrometric unknowns of a single star, this implying that the astrometric part of the design matrix of the system can be easily arranged in a block-diagonal form. This is the basic assumption of the current parallelization schema \cite{Becciani2014}. An arc-based model, with the involvement of two stars at the same time, completely disrupts this structure, so a different approach is needed. A possible solution, that minimizes the communication among the various Processing Elements, is based on grouping the observations on a combination of source and time parameters, and is currently under investigation.

\section{Conclusions}
The Gaia mission represents a great challenge from many different points of view. In particular, its success depends on the correct solution of a linearized system of equations, which enables the definition of a global celestial reference system and the estimation of absolute parallaxes, with profound consequences on all fields of astronomy and beyond. Simply because of its sheer size, this problem, dubbed Global Astrometric Sphere Reconstruction, is a challenging HPC task which, in the Gaia mission, has been tackled in two different ways. The very nature of the Gaia approach, and to some extent of any Gaia-like mission, however, inevitably introduces some strong correlations among the different unknowns of the problem, which can hamper the accuracy of the final catalog. Although the most delicate issues are likely to originate from uncalibrated instrument effects, these might be influenced also by the correlations with the attitude of the satellite. An alternative astrometric model based on the arcs between two stars can help to mitigate these problems and is currently under study. This, however, will require a new parallelization schema, possibly based on grouping observations both by sources and time.

\begin{credits}
\subsubsection{\ackname} This work was supported by the Agenzia Spaziale Italiana (ASI) through contract No. 2018-24-HH.0 and its addendum No. 2018-24-HH.1-2022 to the Italian Istituto Nazionale di Astrofisica (INAF).

\end{credits}
\bibliographystyle{splncs04}
\bibliography{biblio}

\begin{thebibliography}{10}
\providecommand{\url}[1]{\texttt{#1}}
\providecommand{\urlprefix}{URL }
\providecommand{\doi}[1]{https://doi.org/#1}

\bibitem{Becciani2014}
Becciani, U., Sciacca, E., Bandieramonte, M., Vecchiato, A., Bucciarelli, B.,
  Lattanzi, M.G.: Solving a very large-scale sparse linear system with a
  parallel algorithm in the gaia mission. In: 2014 International Conference on
  High Performance Computing Simulation (HPCS). pp. 104--111 (July 2014).
  \doi{10.1109/HPCSim.2014.6903675}

\bibitem{2024SPIE13101E..1OC}
{Cesare}, V., {Becciani}, U., {Vecchiato}, A., {Lattanzi}, M.G., {Aldinucci},
  M., {Bucciarelli}, B.: {The Gaia AVU-GSR solver: CPU+GPU parallel solutions
  for linear systems solving and covariances calculation toward exascale
  systems}. In: {Ibsen}, J., {Chiozzi}, G. (eds.) Society of Photo-Optical
  Instrumentation Engineers (SPIE) Conference Series. Society of Photo-Optical
  Instrumentation Engineers (SPIE) Conference Series, vol. 13101, p. 131011O
  (Jul 2024). \doi{10.1117/12.3018102}

\bibitem{2016A&A...595A...1G}
{Gaia Collaboration}, {Prusti}, T., {de Bruijne}, J.H.J., {Brown}, A.G.A.,
  {Vallenari}, A., {Babusiaux}, C., {Bailer-Jones}, C.A.L., {Bastian}, U.,
  {Biermann}, M., {Evans}, D.W., et~al.: {The Gaia mission}. Astron.\
  Astrophys.  \textbf{595}, ~A1 (Nov 2016). \doi{10.1051/0004-6361/201629272}

\bibitem{2005USNOC.179.....K}
{Kaplan}, G.H.: {The IAU resolutions on astronomical reference systems, time
  scales, and earth rotation models : explanation and implementation}. Tech.
  rep., U.S. Naval Observatory (2005)

\bibitem{2002moas.book.....K}
{Kovalevsky}, J.: {Modern Astrometry}. {Springer-Verlag Berlin Heidelberg}
  (2002)

\bibitem{2021A&A...649A...4L}
{Lindegren}, L., {Bastian}, U., {Biermann}, M., {Bombrun}, A., {de Torres}, A.,
  {Gerlach}, E., {Geyer}, R., {Hern{\'a}ndez}, J., {Hilger}, T., {Hobbs}, D.,
  {Klioner}, S.A., {Lammers}, U., {McMillan}, P.J., {Ramos-Lerate}, M.,
  {Steidelm{\"u}ller}, H., {Stephenson}, C.A., {van Leeuwen}, F.: {Gaia Early
  Data Release 3. Parallax bias versus magnitude, colour, and position}.
  Astron.\ Astrophys.  \textbf{649}, A4 (May 2021).
  \doi{10.1051/0004-6361/202039653}

\bibitem{2021A&A...649A...2L}
{Lindegren}, L., {Klioner}, S.A., {Hern{\'a}ndez}, J., {Bombrun}, A.,
  {Ramos-Lerate}, M., {Steidelm{\"u}ller}, H., {Bastian}, U., {Biermann}, M.,
  {de Torres}, A., {Gerlach}, E., {Geyer}, R., {Hilger}, T., {Hobbs}, D.,
  {Lammers}, U., {McMillan}, P.J., {Stephenson}, C.A., {Casta{\~n}eda}, J.,
  {Davidson}, M., {Fabricius}, C., {Gracia-Abril}, G., {Portell}, J., {Rowell},
  N., {Teyssier}, D., {Torra}, F., {Bartolom{\'e}}, S., {Clotet}, M.,
  {Garralda}, N., {Gonz{\'a}lez-Vidal}, J.J., {Torra}, J., {Abbas}, U.,
  {Altmann}, M., {Anglada Varela}, E., {Balaguer-N{\'u}{\~n}ez}, L., {Balog},
  Z., {Barache}, C., {Becciani}, U., {Bernet}, M., {Bertone}, S., {Bianchi},
  L., {Bouquillon}, S., {Brown}, A.G.A., {Bucciarelli}, B., {Busonero}, D.,
  {Butkevich}, A.G., {Buzzi}, R., {Cancelliere}, R., {Carlucci}, T., {Charlot},
  P., {Cioni}, M.R.L., {Crosta}, M., {Crowley}, C., {del Peloso}, E.F., {del
  Pozo}, E., {Drimmel}, R., {Esquej}, P., {Fienga}, A., {Fraile}, E., {Gai},
  M., {Garcia-Reinaldos}, M., {Guerra}, R., {Hambly}, N.C., {Hauser}, M.,
  {Jan{\ss}en}, K., {Jordan}, S., {Kostrzewa-Rutkowska}, Z., {Lattanzi}, M.G.,
  {Liao}, S., {Licata}, E., {Lister}, T.A., {L{\"o}ffler}, W., {Marchant},
  J.M., {Masip}, A., {Mignard}, F., {Mints}, A., {Molina}, D., {Mora}, A.,
  {Morbidelli}, R., {Murphy}, C.P., {Pagani}, C., {Panuzzo}, P., {Pe{\~n}alosa
  Esteller}, X., {Poggio}, E., {Re Fiorentin}, P., {Riva}, A., {Sagrist{\`a}
  Sell{\'e}s}, A., {Sanchez Gimenez}, V., {Sarasso}, M., {Sciacca}, E.,
  {Siddiqui}, H.I., {Smart}, R.L., {Souami}, D., {Spagna}, A., {Steele}, I.A.,
  {Taris}, F., {Utrilla}, E., {van Reeven}, W., {Vecchiato}, A.: {Gaia Early
  Data Release 3. The astrometric solution}. Astron.\ Astrophys.  \textbf{649},
  A2 (May 2021). \doi{10.1051/0004-6361/202039709}

\bibitem{2012A&A...538A..78L}
{Lindegren}, L., {Lammers}, U., {Hobbs}, D., {O'Mullane}, W., {Bastian}, U.,
  {Hern{\'a}ndez}, J.: {The astrometric core solution for the Gaia mission.
  Overview of models, algorithms, and software implementation}. Astron.\
  Astrophys.  \textbf{538}, A78 (Feb 2012). \doi{10.1051/0004-6361/201117905}

\bibitem{PaigeSaunders1982}
Paige, C.C., Saunders, M.A.: Lsqr: An algorithm for sparse linear equations and
  sparse least squares. ACM Trans. Math. Softw.  \textbf{8}(1),  43--71 (Mar
  1982). \doi{10.1145/355984.355989}

\bibitem{2014RMxAC..45...35R}
{Riva}, A., {Gai}, M., {Lattanzi}, M.G., {Russo}, F., {Buzzi}, R.: {BAM: A
  metrology device for a high precision astrometric mission}. In: Revista
  Mexicana de Astronomia y Astrofisica Conference Series. Revista Mexicana de
  Astronomia y Astrofisica, vol.~27, vol.~45, pp. 35--38 (Dec 2014)

\bibitem{2018A&A...620A..40V}
{Vecchiato}, A., {Bucciarelli}, B., {Lattanzi}, M.G., {Becciani}, U.,
  {Bianchi}, L., {Abbas}, U., {Sciacca}, E., {Messineo}, R., {De March}, R.:
  {The global sphere reconstruction (GSR). Demonstrating an independent
  implementation of the astrometric core solution for Gaia}. Astron.\
  Astrophys.  \textbf{620}, A40 (Nov 2018). \doi{10.1051/0004-6361/201833254}

\end{thebibliography}

\end{document}